\documentclass{article}

\usepackage{amssymb}
\usepackage{amsmath}
\usepackage{hyperref}
\usepackage[mathscr]{eucal}
\usepackage{authblk}

\title{Sizing the bets in a focused portfolio}

\author[$\dagger$]{Vuko Vuk\v{c}evi\'{c}}
\author[$\dagger$]{Robert Keser}

\affil[$\dagger$]{In silico Ltd., Zagreb, Croatia,\linebreak vuko.vukcevic@insilico.hr, robert.keser@insilico.hr}

\begin{document}

\maketitle

\section{Summary}
\label{sec:summary}

The paper provides a mathematical model and a tool for the focused
investing strategy as advocated by Buffett~\cite{berkshireLetters},
Munger~\cite{almanack}, and others from this investment community.
The approach presented here assumes that the investor's role is to think about
probabilities of different outcomes for a set of businesses. Based on these
assumptions, the tool calculates the optimal allocation of capital for
each of the investment candidates. The model is based on a generalized Kelly
Criterion with options to provide constraints that ensure: no shorting, limited
use of leverage, providing a maximum limit to the risk of permanent loss of
capital, and maximum individual allocation. The software is applied to an
example portfolio from which certain observations about excessive
diversification are obtained. In addition, the software is made available for
public use.\\

\section{Introduction}
\label{sec:introduction}

A stock is an ownership share of business~\cite{intelligentInvestor}, in
which time--scales are significantly greater than the time--scales of
day--to--day stock price fluctuations. The work therefore assumes a
fundamentals--based investment analysis with a very long time horizon. There are
generally two parts of a fundamentals--based investment process:
\begin{enumerate}
    \item Finding businesses that are investment candidates,
    \item Deciding how much capital to allocate in each of them.
\end{enumerate}

\noindent This work does not deal with the first part related to finding
relevant investment candidates. It deals only with the second part of the
process, which can be mathematically well formulated, in an attempt to minimize
the effects of the psychological misjudgements such as anchoring
bias~\cite{kahneman}, consistency and commitment tendency~\cite{cialdini}, and
others with their combined effects~\cite{almanack}). Essentially, the question
to answer is: For a set of candidate companies and their current market
capitalization, each having a set of scenarios defined by a probability and
intrinsic value estimate, how much of capital to invest in each?\\

\indent The answer to this question has been given by Kelly~\cite{kelly} with
his widely known Kelly Criterion. Despite the fact that the Kelly Criterion has
been published in 1956 and dealt with information theory and horse racing, its
use is now widespread in the field of finance, with a plethora of recent
research. For example, Pinelis and Ruppret~\cite{pinelisRuppert} noted that the
asset allocation fractions (weights) of the machine--learning powered Maximally
Predictable Portfolio (MPP) are similar to weighting by the Kelly Criterion.
Meister~\cite{meister} used the Kelly Criterion to demonstrate the advantage of
an investor who bets on multiple simultaneous investments compared to an
investor who bets on a single investment. In the limit of infinite number of
simultaneous and similar bets, the first investor has all the capital invested
in a diversified fashion, providing a clear advantage. Byrnes and
Barnett~\cite{byrnesBarnett} went further to generalize the Kelly Criterion with
a continuous probability distribution modeling the price movement of the stock,
while Lototsky and Pollok~\cite{lototskyPollok} rigorously analyzed the
high--frequency limit of the model, concluding that a high--frequency
strategy can lead to a more aggressive betting strategy.  Proskurnikov and
Barmish~\cite{proskurnikovBarmish} investigated the possibility of having a
non--linear control for the Kelly betting, assuming a bounded set for the
input probabilities instead of assuming that they are known in advance.  The
Kelly Criterion was recently even extended to the quantum realm by Meister
and Price~\cite{meisterPrice}, where in addition to the uncertainty related
to flipping a coin, there is an uncertainty in measurement. Perhaps the most
related reference to this work is the paper by Jim\'{e}nez et
al.~\cite{jimenezEtAl} where they extended the Kelly Criterion to both
multivariate and simultaneous bets and applied it to sports betting in the
English Premier League. This work largely follows the approach by
Jim\'{e}nez et al.~\cite{jimenezEtAl}, but applied to a fundamentals based,
focused investing with the following characteristics:
\begin{enumerate}
    \item Multiple companies (stocks) are considered simultaneously,
    \item Each with an arbitrary number of scenarios,
    \item Multiple constraints for modeling authors' preference towards no
    shorting, limited use of leverage, and providing a maximum value for the
    risk of permanent loss of capital,
    \item A fundamentals--based analysis with a very long time horizon, without
    necessarily knowing when the market price will correspond to the
    fundamentals.
\end{enumerate}

The generalization leads to a non--linear system of equations that when solved,
yields a fraction of capital to invest in each of the candidate companies. The
mathematical derivation is presented in \autoref{sec:mathematics}, while in the
next two subsections (\autoref{sec:assumptions} and \autoref{sec:safetyMargin}),
a discussion on assumptions and the margin of safety, are presented.

\subsection{Assumptions}
\label{sec:assumptions}

The underlying philosophy with the fundamentals--based focused investing is that
one should spend the majority of their time analyzing investments and thinking
about intrinsic values under different scenarios that might play out. However,
calculating intrinsic value of a company is a process subject to multiple levels
of uncertainty. Therefore, the reader should always keep in mind that if the
inputs to the mathematical model are unrealistic, the results will be
unrealistic as well.

During the mathematical derivation presented in \autoref{sec:mathematics}, an
assumption is made that the number of bets is very high (tends to infinity).
The assumption is made in order to write the equations in terms of probabilities
instead of the number of outcomes divided by the number of bets. Therefore, as
long as the input probabilities are \textit{conservatively} estimated, the
framework should still be valid. This essentially represents the most important
margin of safety~\cite{intelligentInvestor}.

\subsection{Margin of Safety}
\label{sec:safetyMargin}

In addition to the most important margin of safety mentioned above, there are
however a couple of more margins of safety embedded in the current framework.
These are:
\begin{itemize}
    \item No shorting allowed. From a purely mathematical point of view,
    shorting would be allowed. Without detailed mathematics, it is easy to see
    how a company with a negative expected return would result in a short
    position. However, due to the asymmetry of the potential losses compared to
    gains, coupled with the usual time--frame limit that comes with shorting, we
    provided a constraint for being long--only.
    \item Limited use of leverage. Although leverage is especially useful when
    dealing with high--conviction ideas, this constraint is introduced to model
    authors' conservative behavior.
    \item No company without at least one downside scenario is allowed. By
    disallowing inputs without downside, the framework forces us to focus and
    think about what can go wrong, as opposed to focusing on what can go
    right. This is in--line with thoughts by Prasad~\cite{darwinInvesting} about
    the importance of avoiding the errors of investing in a bad company.
    However, if one is absolutely (100\%) sure that a company does not have a
    downside, the solution is to put a significant amount of leveraged capital
    in that one company. There are two options to handle such cases: i) Either
    outside of this framework or ii) Modeling an unknown downside scenario with
    a small probability and intrinsic value of zero.
\end{itemize}

These assumptions and margins of safety are embedded into the framework as
constraints in order to try and tie the mathematical model and the uncertainty
around inputs.

\section{Mathematics}
\label{sec:mathematics}

The following derivation mostly follows the first part of the work by Byrnes and
Barnett~\cite{byrnesBarnett}. The problem statement is repeated here for
convenience: Given a set of candidate companies, each having a set of scenarios
described by the probability $p$ and estimate of the intrinsic value
$\mathcal{V}$, calculate the optimal allocation fraction $f$ for each candidate
company by maximizing the long--term growth rate of assets. After a single
outcome (realization), the change in value of assets can be written as follows:
\begin{equation}
\label{eq:1}
    \mathcal{A}_{after}
  = 
    \mathcal{A}_{before}
    \left( 1 + \sum_{j}^{N_c} f_j k_j \right)
\end{equation} 

\noindent where $\mathcal{A}$ is the value of assets (capital), $N_c$ is the
number of candidate companies to consider, $f_j$ is the allocated fraction to
$j$th company, and $k_j$ is a return for a company $j$ defined as the relative
difference between the estimated intrinsic value under a given scenario
($\mathcal{V}$) and the market capitalization at the time of investing
($\mathcal{M}$):
\begin{equation}
\label{eq:2}
    k_j = \frac{\mathcal{V}_j - \mathcal{M}_j}{\mathcal{M}_j}
\end{equation}

\noindent If a significant number of (re)allocations ($N_a$) is performed in
succession, the equation~\eqref{eq:1} can be written as follows:
\begin{equation}
\label{eq:3}
    \mathcal{A}_{N_a}
  = 
    \mathcal{A}_{0} \prod_{i_1, i_2, \hdots, i_{N_o}}
    \left( 1 + \sum_{j}^{N_c} f_j k_{ij} \right)^{n_i}
\end{equation} 

\noindent where $n_i \in {n_{i1}, n_{i2}, \hdots, n_{N_o}}$ is the number of
times the $i$th outcome has occurred. Note that $k_{ij}$ represents the return
of the $j$th company for the $i$th outcome. Following original Kelly's approach,
a logarithmic growth function $\mathcal{G}$ is introduced:
\begin{equation}
\label{eq:4}
    \mathcal{G} = \lim_{N_a \to \infty} \frac{1}{N_a} \ln
        \frac{\mathcal{A}_{N_a}}{\mathcal{A}_0}
\end{equation}

\noindent and the goal is to find its maximum with respect to allocation
fractions $f_j$:
\begin{equation}
\label{eq:5}
    \frac{\partial \mathcal{G}}{\partial f_j} = 0
\end{equation}

\noindent Substituting equation~\eqref{eq:1} into equation~\eqref{eq:5} results
in the following equation, after some calculus and algebra:
\begin{equation}
\label{eq:6}
    \lim_{N_a \to \infty} \frac{1}{N_a}
    \sum_{i}^{N_o} \frac{n_i k_{ij}}{1 + \sum_{j}^{N_c} f_j k_{ij}} = 0
\end{equation}

\noindent If one assumes an infinite number of (re)allocations
$N_a$\footnote{The assumption regarding the infinite number of allocations is
somewhat incompatible with a focused investment strategy where a small number of
stocks are considered at any point in time. This incompatibility is addressed by
margins of safety presented in~\autoref{sec:safetyMargin}.}, the following
relation holds:
\begin{equation}
\label{eq:7}
    \lim_{N_a \to \infty} \frac{n_i}{N} = p_i
\end{equation}

\noindent Where $p_i$ is the probability of the $i$th outcome. For example, if
there are two companies, each with two 50--50 scenarios, there will be four
outcomes in total with the probability of each outcome equal to 25\%. Finally,
substituting equation~\eqref{eq:7} into~\eqref{eq:6} results in a system of
equations written in terms of probabilities $p_i$, expected returns for each
company in each of the outcomes $k_{ij}$, and allocation fractions for each
company $f_j$:
\begin{equation}
\label{eq:8}
    \sum_{i}^{N_o} \frac{p_i k_{ij}}{1 + \sum_{j}^{N_c} f_j k_{ij}} = 0
\end{equation}

\noindent The equation~\eqref{eq:8} represents a non--linear system of equations
in the unknown fractions $f_j$, which when solved, should yield optimal
allocation strategy for maximizing long--term growth of capital.

\subsection{Constraints}
\label{sec:constraints}

\noindent Equation~\eqref{eq:8} has no constraints, meaning that after solving
the system of equations, the resulting fractions $f_j$ might be negative and
greater than one. This would imply short positions and use of leverage,
respectively. A general inequality constraint may be written in the following
form:
\begin{equation}
\label{eq:9}
    I(f_j) \le 0
\end{equation}

\noindent where, for example, $I(f_j) \equiv -f \le 0$ models a long-only
constraint that would make sure that the fractions are positive. In order to
transform the inequality constraint into an equality constraint, we introduce a
slack variable $s$ that must be positive:
\begin{equation}
\label{eq:10}
    I(f_j) + s = 0 \equiv \mathcal{C}(f_j, s) = 0
\end{equation}

\noindent where the second part of the equation introduces a useful substitution
for deriving the constrained system later on.\\

\indent In this work, we define four constraints that serve as additional
margins of safety in a focused investment approach:
\begin{enumerate}
    \item Long--only constraint ensures that a fraction cannot be negative and
    is added to all candidate companies. It disallows short positions:
    \begin{equation}
        \label{eq:11}
            f \ge 0 \rightarrow -f + s = 0
        \end{equation}

    \item Maximum leverage constraint ensures that the leverage is limited up to
    $L$. Note that $L = 0$ implies no leverage:
        \begin{equation}
        \label{eq:12}
            \sum_j^{N_c} f_j \le 1 + L \rightarrow \sum_j^{N_c} f_j - 1 - L + s = 0
        \end{equation}

    \item Maximum individual allocation constraint ensures that a fraction does
    not exceed the specified amount. Adding this constraint allows preventing
    excessive concentration in the portfolio:
        \begin{equation}
        \label{eq:13}
            f \le M \rightarrow f - M + s = 0
        \end{equation}

    \item Maximum allowable permanent capital loss constraint ensures that the
    worst--case outcome does not exceed losing a specified amount of capital
    with a specified probability:
        \begin{equation}
        \label{eq:14}
            \sum_j^{N_c} f_j \min(p_{ij} k_{ij}) \ge P \cdot K \rightarrow
            -\sum_j^{N_c} \left(f_j \min(p_{ij} k_{ij}) \right) + P \cdot K + s = 0
        \end{equation}
    \noindent where $\min(p_{ij} k_{ij})$ is the worst--case outcome across all
    scenarios for the $j$-th candidate company. Here, the minimum returns
    $k_{ij}$ and the maximum worst--case return $K$ are both negative by
    convention, indicating a loss, hence the $\ge$ sign.
\end{enumerate}

\indent Note that the maximum allowable permanent capital loss constraint
given by equation~\eqref{eq:14} only works with the long--only constraint
because the fraction $f_j$ is assumed positive. It is also important to note
that here we do not refer to a {\it temporary} loss of capital due to short--term
stock market fluctuations, but rather {\it permanent} loss of capital due to the
fundamental business environment of candidate companies. In addition, it is of
course possible for one to lose more than the specified maximum allowable amount
of capital because of the assumptions made with respect to the inputs.

\subsection{Constrained System}
\label{sec:constrainedSystem}

\noindent The growth function~\eqref{eq:4} can be constrained with an arbitrary
amount of constraints~\eqref{eq:10} by introducing a Lagrangian:
\begin{equation}
\label{eq:15}
    \mathcal{L}(f_j, \lambda_l)
  =
    \mathcal{G}(f_j) - \sum_l^{N_l} \lambda_l \mathcal{C}_l(f_j, s_l)
\end{equation}

\noindent where $N_l$ is the number of constraints and $l$ denotes the $l$-th
constraint defined either by $l$-th Lagrange multiplier $\lambda_l$ or the
$l$-th slack variable $s_l$. There are two necessary conditions for finding a
constrained maximum of the growth function:
\begin{equation}
\label{eq:16}
    \alpha_i(f_j, \lambda_l, s_l)
  =
    \frac{\partial \mathcal{L}(f_j, \lambda_l)}{\partial f_j}
  =
    \frac{\partial \mathcal{G}(f_j)}{\partial f_j}
  - \sum_l^{N_l} \lambda_l \frac{\partial{\mathcal{C}_l(f_j, s_l)}}{\partial f_j}
  =
    0
\end{equation}

\begin{equation}
\label{eq:17}
    \beta_i(f_j, s_l)
  =
    \frac{\partial \mathcal{L}(f_j, \lambda_l)}{\partial \lambda_l}
  =
    - \mathcal{C}_l(f_j, s_l)
  =
    0
\end{equation}

\noindent where $\alpha_i$ and $\beta_i$ have been introduced as a shorthand
notation that distinguishes between two vector equations: There are $N_c$
of $\alpha$ equations and $N_l$ of $\beta$ equations. Therefore, there are
$N_c + N_l$ equations, but $N_c + 2N_l$ unknowns, because of $N_c$ unknown
fractions, $N_l$ unknown Lagrange multipliers $\lambda_l$ and $N_l$ unknown
slack variables $s_l$. However, an inequality constraint cannot be active and
inactive at the same time. An active constraint is characterized by $\lambda_l
\neq 0$ and $s_l = 0$, whereas an inactive constraint is characterized by
$\lambda_l = 0$ and $s_l > 0 \ne 0$. This means that we have to solve $2^{N_l}$
nonlinear systems to cover all combinations of constraints and pick the best
solution. As an example, adding all constraints mentioned
in~\autoref{sec:constraints} for a portfolio with five candidate companies would
result in having to solve $2^{12} = 4 096$ systems, while having ten candidate
companies would imply solving $2^{22} = 4 194 304$ systems, demonstrating the
exponential complexity of the problem.

\section{Numerics}
\label{sec:numerics}

\noindent The equations~\eqref{eq:16} and~\eqref{eq:17} can be written
succinctly as:
\begin{equation}
\label{eq:18}
    \mathcal{F}_i(x_i) = 0
\end{equation}

\noindent where $x_i$ is a vector of unknown fractions and unknown Lagrange
multipliers or slack variables:
\begin{equation}
\label{eq:19}
    x_i
  =
    \{ f_1, f_2, \ldots, f_{N_c}, \lambda_1 | s_1, \lambda_2 | s_2,
    \ldots, \lambda_{N_l} | s_{N_l} \}
\end{equation}

\indent Because $\mathcal{F}_i$ is a non--linear equation in $f_j$,
the Newton--Raphson method is used to find a numerical solution. The method is
iterative and starts by linearizing the equation around the previous solution
from the previous iteration:
\begin{equation}
\label{eq:20}
    \mathcal{F}_i^o + \sum_{i}^{N_c + N_l} \mathcal{J}_{ij}^o (x_j^n - x_j^o) = 0
\end{equation}

\noindent where $\mathcal{J}_{ij}$ is the Jacobian of $\mathcal{F}_i$, and
superscripts $^n$ and $^o$ denote the new value and the old value, respectively.

\subsection{Jacobian for an Active Constraint}
\label{sec:jacobianActive}

\noindent If a constraint is active ($\lambda_l \ne 0, s_l = 0$), the
corresponding unknown is the Lagrange multiplier and the Jacobian has the
following form:
\begin{equation}
\label{eq:21}
    \mathcal{J}_{ij}
  =
    \begin{pmatrix}
        \frac{\partial \alpha_i}{\partial f_j}
      & \frac{\partial \alpha_i}{\partial \lambda_j} \\
        \frac{\partial \beta_i}{\partial f_j}
      & \frac{\partial \beta_i}{\partial \lambda_j} \\
    \end{pmatrix}
  =
    \begin{pmatrix}
        \frac{\partial^2 \mathcal{G}}{\partial f_i \partial f_j}
      - \sum_j^{N_l} \left( \lambda_j \frac{\partial^2 \mathcal{C}_i}{\partial f_j^2} \right)
      & -\frac{\partial \mathcal{C}_i}{\partial f_j} \\
        -\frac{\partial \mathcal{C}_i}{\partial f_j}
      & 0 \\
    \end{pmatrix}
\end{equation}

\subsection{Jacobian for an Inactive Constraint}
\label{sec:jacobianInactive}

\noindent If a constraint is inactive ($\lambda_l = 0, s_l > 0 \ne 0$), the
corresponding unknown is the slack variable and the Jacobian has the following
form:
\begin{equation}
\label{eq:22}
    \mathcal{J}_{ij}
  =
    \begin{pmatrix}
        \frac{\partial \alpha_i}{\partial f_j}
      & \frac{\partial \alpha_i}{\partial s_j} \\
        \frac{\partial \beta_i}{\partial f_j}
      & \frac{\partial \beta_i}{\partial s_j} \\
    \end{pmatrix}
  =
    \begin{pmatrix}
        \frac{\partial^2 \mathcal{G}}{\partial f_i \partial f_j}
      & 0 \\
        -\frac{\partial \mathcal{C}_i}{\partial f_j}
      & -1 \\
    \end{pmatrix}
\end{equation}

\noindent Partial derivatives of the constraint function can be readily obtained
because all constraints presented in this work are simple analytical functions
(see~\autoref{sec:constraints}). For completeness, the Hessian of the growth
function that appears in the upper--right corner is:
\begin{equation}
\label{eq:23}
    \frac{\partial^2 \mathcal{G}}{\partial f_i \partial f_j}
  = 
    - \sum_{o}^{N_o}
      \frac{p_o k_{oi} k_{oj}}{\left(1 + \sum_{j}^{N_c} f_j k_{ij} \right)^2}
\end{equation}

\noindent where subscript for outcome $i$ has been changed to $o$ in order to be
able to write the Hessian in the standard $ij$ notation.

\indent Note that the equations~\eqref{eq:21} and~\eqref{eq:22} are presented in
a way that all constraints are either active or inactive. Since we have to solve
$2^{N_l}$ systems characterized by active/inactive status of each constraint,
that simply means that we need to add respective active/inactive contributions
to the Jacobian $\mathcal{J}_{ij}$ and the right--hand--side $\mathcal{F}_i^o$
in equation~\eqref{eq:20} for a particular constraint $j$.

\subsection{Notes on the Numerical Procedure}
\label{sec:numericalNotes}

\noindent The numerical procedure starts by assuming the uniform allocation
across all companies, i.e. $f_j = f = \frac{1}{N_c}$. Based on the $f_j^o$ in
the current iteration, the linear system in equation~\eqref{eq:10} is solved to
find the new solution $f_j^n$. The process is repeated until sufficient level of
accuracy is reached.\\

\indent After solving all $2^N_l$ solutions, we end up with less than $2^N_l$
viable solutions. A solution is considered viable if the Newton--Raphson
procedure managed to find a numerical solution and if all slack variables
for all inactive constraints are positive. Finding the best solution out of all
viable solutions would require evaluating the growth function for all
solutions, which is particularly challenging due to the product series
in equation~\eqref{eq:3}. In this work, we take a pragmatic approach and
simply choose the solution that maximizes the expected value of the portfolio
out of a set of most diversified solutions, i.e. the solutions with the highest
number of non--zero allocations.

\section{Basic Validation Tests}
\label{sec:validation}

\noindent In order to validate the numerical model, a basic example of five
candidate companies is considered, where each candidate has the same set of
scenarios (probabilities and intrinsic values) and the same market cap.
The inputs that represent a 50\% loss and 100\% gain with 50--50 probabilities
are presented in~\autoref{tab:validationCompany}.\\

\begin{table}[!ht]
\caption{Company for a validation test with a market cap of 1.}
\vspace{0.25cm}
\centering
\begin{tabular}{l|c|c}
Scenario & Intrinsic value & Probability \\
\hline
50\% down with 50\% probability & 0.5 & 50\% \\
100\% up with 50\% probability & 2 & 50\% \\
\end{tabular}%
\label{tab:validationCompany}%
\end{table}%

\indent Solving the system without any constraints yields a uniform allocation
of 35\% of capital in each company. Note that because we considered 5 companies,
that implies 75\% leverage ($5 \cdot 35\% = 175\%$). Even without the
constraint for maximum allocation of capital, just maximizing the long--term
growth--rate of assets prefers a diversified solution, which is expected.\\

\indent Adding a constraint for no leverage (given by equation~\eqref{eq:12} and
setting $L = 0$), results in a uniform allocation of 20\%, as expected. It is
straightforward to show that the worst--case outcome in such a portfolio implies
permanently losing 50\% of the capital with probability of 3.125\%.\\

\indent The final test is done by setting the maximum allowable permanent
capital loss constraint as given by equation~\eqref{eq:14}. Setting $P = 5\%$
and $K = 50\%$, indicating that we are comfortable risking to lose 50\% of the
capital with 5\% probability, results in a uniform allocation of 2\%. Because
only 10\% of capital is invested in that case, there is a possibility of losing
5\% of capital with probability of 3.125\%. In the worst--case scenario, the
probability--weighted return is $-0.5 \cdot 0.02 \cdot 0.5 \cdot 5 = -2.5\%$,
which is equal to $P \cdot K$.

\section{Example}
\label{sec:example}

\noindent With the basic validation of the numerical model performed, the
attention is moved to a realistic example. Consider five candidate companies,
each with up to three scenarios. Each scenario is represented by an intrinsic
value and the probability of the scenario happening (or intrinsic value being
reached at some point in the future). Note that how these numbers are obtained
is outside of the scope of this work, although it is important to stress that
the validity and conservative assumptions behind these numbers are probably the
most important part of an investor's job who is taking a fundamental analysis
approach. The example inputs are presented in~\autoref{tab:companyA}
to~\autoref{tab:companyE}.

\begin{table}[!b]
\caption{Company A with current market cap of 225B USD.}
\vspace{0.25cm}
\centering
\begin{tabular}{l|c|c}
Scenario & Intrinsic value & Probability \\
\hline
Total loss & 0 USD & 5\% \\
Base thesis & 270B USD & 60\% \\
Bull thesis & 420B USD & 35\% \\
\end{tabular}%
\label{tab:companyA}%
\end{table}%

\begin{table}[!b]
\caption{Company B with current market cap of 450M USD.}
\vspace{0.25cm}
\centering
\begin{tabular}{l|c|c}
Scenario & Intrinsic value & Probability \\
\hline
Total loss& 0 USD & 5\% \\
Bear thesis & 350M USD & 50\% \\
Base thesis & 900M USD & 45\% \\
\end{tabular}%
\label{tab:companyB}%
\end{table}%

\begin{table}[!b]
\caption{Company C with current market cap of 39M GBP.}
\vspace{0.25cm}
\centering
\begin{tabular}{l|c|c}
Scenario & Intrinsic value & Probability \\
\hline
Total loss & 0 GBP & 10\% \\
Bear thesis & 34M GBP & 40\% \\
Base thesis & 135M GBP & 50\% \\
\end{tabular}%
\label{tab:companyC}%
\end{table}%

\begin{table}[!b]
\caption{Company D with current market cap of 751M SGD.}
\vspace{0.25cm}
\centering
\begin{tabular}{l|c|c}
Scenario & Intrinsic value & Probability \\
\hline
Bear thesis & 330M SGD & 30\% \\
Base thesis & 1B SGD & 70\% \\
\end{tabular}%
\label{tab:companyD}%
\end{table}%

\begin{table}[!b]
\caption{Company E with current market cap of 126B HKD.}
\vspace{0.25cm}
\centering
\begin{tabular}{l|c|c}
Scenario & Intrinsic value & Probability \\
\hline
Total loss & 0 HKD & 5\% \\
Bear thesis & 50B HKD & 10\% \\
Base thesis & 300B HKD & 85\% \\
\end{tabular}%
\label{tab:companyE}%
\end{table}%

\indent Based on these inputs, with the long--only strategy, without leverage,
and with maximum individual allocation of 30\%, the portfolio allocation that
maximizes the long--term growth--rate of capital is presented
in~\autoref{tab:results}.

\begin{table}
\caption{Portfolio that maximizes long--term growth--rate of capital.}
\vspace{0.25cm}
\centering
\begin{tabular}{l|c|c|c|c|c|c}
Company & A & B & C & D & E \\
\hline
Allocation fractions & 30\% & 8\% & 30\% & 2\% & 30\% \\
\end{tabular}%
\label{tab:results}%
\end{table}%

With the obtained fractions, it is easy to obtain some useful
statistics on the portfolio, namely:
\begin{itemize}
    \item Expected gain of 32 cents for every dollar invested,
    \item Cumulative probability of loss of capital of 16\%,
    \item Permanent loss of 60\% of capital with probability of 0.008\%.
\end{itemize}

\noindent The probability of permanent loss of capital of 0.008\% is
particularly interesting. Provided that the inputs are reasonably estimated, and
considering that the portfolio has five stocks, that is a very strong argument
against excessive diversification, especially if:
\begin{itemize}
    \item One thinks of stocks as ownership shares of businesses, which implies
    long--term thinking and not being bothered by market fluctuations, 
    \item One embeds a margin of safety in different scenarios for different
    companies by e.g. recognizing that both unknown and known bad things may
    happen.
\end{itemize}

\noindent The observation about excessive diversification is in--line with the
thoughts from the Poor Charlie's Almanack~\cite{almanack} and one of the
lectures from Li Lu~\cite{liLuLecture}. In addition, it seems that recent
research is also drawing similar conclusions about excessive
diversification~\cite{jamesMenzies,arvanitisEtAl}, although using different
investing frameworks. The study by James and Menzies~\cite{jamesMenzies} is
particularly interesting to the authors since the data suggests that strong
investors should hold concentrated portfolios, which is in--line with practical
observations of extremely successful fundamentals--based investors: Warren
Buffett~\cite{berkshireLetters}, Mohnish Pabrai, Nick Sleep~\cite{nomadLetters},
Pulak Prasad~\cite{darwinInvesting}, and others.

\section{Problems, Discussion, and Future Work}
\label{sec:futureWork}

\noindent A couple of problems have been observed by the authors:
\begin{enumerate}
    \item It is possible that a given non--linear system for a particular
    combination of constraint statuses (active/inactive) does not converge. This
    may happen if the resulting matrix is singular, or if the Newton--Raphson
    algorithm does not find the solution within a prescribed number of
    iterations. These errors are ignored, which means that it is possible that
    the optimal solution is not found due to numerical issues.
    \item The exponential complexity of the model makes it challenging to
    use for more than several candidate companies without using significant
    compute resources. For example, having a 20 candidate companies with all
    constraints would result in around 4 trillion non--linear systems to solve.
    Therefore, the model is not suitable for cases with excessive
    diversification, although there is a possibility to filter some of these
    upfront without attempting to solve them, which may be one of the topics for
    future work.
\end{enumerate}

\indent To conclude, we believe that the most challenging aspect of an
investor's work that might use this software is to appropriately and
conservatively model the inputs. The authors see the usefulness of this software
mainly in:
\begin{itemize}
    \item Forcing the investors to think consistently in terms of probabilities
    and long--term business outcomes across the range of candidate companies,
    \item Avoiding psychological biases by having a tool that calculates the
    optimal allocation of capital based on a strict mathematical formalism.
\end{itemize}

\section{Access to the Software}

\noindent The software is freely available as a Rust crate
at~\url{https://crates.io/crates/charlie}, and its source code is hosted on
GitLab at~\url{https://gitlab.com/in-silico-public/charlie}. The reader is
referred to its documentation for instructions on how to install and use the
software.

\clearpage

\bibliographystyle{biolett}
\bibliography{./references}

\end{document}